\documentclass[11pt]{article}
\usepackage{epsfig} 
\newcommand{\sect}[1]{ \section{#1} \setcounter{equation}{0} }

\newcommand{\Dslash}{D \! \! \! \! /}

\newcommand{\half}{\mbox{\small{$\frac{1}{2}$}}} 
 
\newcommand{\Nf}{N_{\!f}} 
\newcommand{\MSbar}{\overline{\mbox{MS}}}

\begin{document}
\title{Three loop $\MSbar$ renormalization of QED in the 't~Hooft-Veltman
gauge} 
\author{J.A. Gracey, \\ Theoretical Physics Division, \\ 
Department of Mathematical Sciences, \\ University of Liverpool, \\ P.O. Box 
147, \\ Liverpool, \\ L69 3BX, \\ United Kingdom.} 
\date{} 
\maketitle 
\vspace{5cm} 
\noindent 
{\bf Abstract.} Quantum electrodynamics (QED) fixed in the 't~Hooft-Veltman 
gauge is renormalized to three loops in the $\MSbar$ scheme. The 
$\beta$-functions and anomalous dimensions are computed as functions of the
usual QED coupling and the additional coupling, $\xi$, which is introduced as 
part of the nonlinear gauge fixing condition. Similar to the maximal abelian 
gauge of quantum chromodynamics, the renormalization of the gauge parameter is 
singular.  

\vspace{-17cm}
\hspace{13.5cm}
{\bf LTH 767}

\newpage 

\sect{Introduction.}
Gauge theories are the underlying quantum field theory describing the physics
of the fundamental properties of nature. For instance, the field theory 
describing the strong interaction, quantum chromodynamics (QCD), is based on a 
non-abelian colour group and exhibits asymptotic freedom, \cite{1,2}, whereby
the basic quarks and gluons effectively behave as free particles at very high 
energies. Indeed this property allows one to apply perturbation theory to 
describe strong physics phenomena such as deep inelastic scattering to very 
high precision. One fundamental reason for the basic feature of asymptotic 
freedom, \cite{1,2}, is the self-interaction of the gluon which is a natural 
consequence of the generalization of the abelian gauge theory of quantum 
electrodynamics (QED) to the non-abelian gauge group. The consequent
nonlinearities introduced in this extension provide a richer structure not
present in the dynamics of electrons and photons. In order to perform quantum
calculations to determine the properties of a gauge theory, one has to first 
fix the gauge to ensure that only the physical degrees of freedom are taken
into account. There is a large range of choices as to how to eliminate the
unphysical degrees of freedom. In general such choices can be classified
roughly under various headings, such as covariant or non-covariant, linear or
nonlinear and physical or unphysical. (For a comprehensive review see, for
example, \cite{3}.) Ordinarily when one wishes to perform high precision and 
therefore high loop calculations, one chooses linear covariant gauges (which 
are not physical) such as the Landau or Feynman gauge. This is primarily 
because their covariant, though unphysical, nature does not overcomplicate the 
resultant (massless) Feynman diagrams. By contrast other gauges, whilst being 
motivated by physical considerations, such as the Coulomb gauge, are not 
necessarily renormalizable. If they can be proved to be renormalizable at a 
formal level, then it is not always clear whether loop integrals beyond one 
loop can be computed, \cite{3}. However, irrespective of how one fixes the 
gauge, one fundamental feature is always present and that is that physical 
predictions must always be independent of the choice of gauge. As the bulk of 
multiloop calculations are performed in linear covariant gauges and mass 
independent renormalization schemes, one check in this instance is that the 
covariant gauge fixing parameter, $\alpha$, must be absent in the determination
of physical quantities or objects which are clearly gauge independent or gauge 
invariant, \cite{4}. Indeed this property provides a powerful checking tool in 
intricate multiloop calculations. 

Although linear covariant gauges have been examined in detail, there has been 
recent interest in {\em nonlinear} covariant gauges due to their potential 
connection with the infrared dynamics of non-abelian gauge theories, 
\cite{5,6}. For instance, the Curci-Ferrari gauge, \cite{7}, and maximal 
abelian gauge (MAG), \cite{8,9,10}, have been studied where the aim was to 
examine the abelian dominance hypothesis, \cite{8,11,12,13}. Briefly the gluons
in the centre of the colour group are believed to dominate the infrared sector 
due to the off-diagonal gluons acquiring a dynamical mass greater than that of 
the centre gluons. The latter are then central to abelian monopole condensation
that is believed to drive the confinement mechanism, \cite{8,11,12,13}. 
However, from a practical point of view to perform any calculations in such 
nonlinear gauges one needs to renormalize the gauge theory to as high a loop 
order as is possible. This has been achieved for both the Curci-Ferrari gauge 
and MAG in the $\MSbar$ scheme at three loops, \cite{14,15}. From the point of 
view of trying to understand basic features of covariant nonlinearly gauge 
fixed gauge theories both these gauges have common properties. For instance, 
unlike linear covariant gauges they have quartic ghost interactions and the 
corresponding gauge parameters get non-trivial renormalization. Further in the 
case of the MAG this gauge parameter renormalization is singular. Though gauge 
independent quantities, such as the $\beta$-function, correctly emerge as gauge
parameter independent. Whilst these gauges are essentially related by 
construction to the non-abelian aspect of the gauge theory itself, one natural 
question to ask is, is this a feature in other nonlinear gauges when one has an
abelian structure.

Clearly QED is invariably treated in a linear covariant gauge. However, with 
the explosion of interest in gauge theories in the early 1970's QED was gauge 
fixed in a nonlinear gauge known as the 't~Hooft-Veltman gauge, \cite{16}, and
shown to be renormalizable. The primary interest in this gauge fixing was that
the nonlinearity naturally introduced an abelian gauge theory which mimicked
QCD. This was due not only to the presence of interacting Faddeev-Popov ghosts 
but also triple and quartic {\em photon} self-interactions. As the latter do 
not appear in linear covariant gauges in QED, the 't~Hooft-Veltman gauge 
clearly could be used as a laboratory to study simple issues related to gauge 
field self-interaction. Indeed in \cite{17,18} a one loop calculation showed 
that the corresponding covariant gauge fixing parameter was renormalized. 
Therefore, in light of these observations it is the purpose of this article to 
record the full three loop renormalization of QED in the 't~Hooft-Veltman 
gauge. Indeed as far as we are aware this will represent the {\em first} 
detailed multiloop study of the renormalization of QED in the 't~Hooft-Veltman 
gauge. We will construct all the renormalization group functions in the 
$\MSbar$ scheme including the $\beta$-function of the electron-photon coupling
constant which will agree with the already established results of 
\cite{19,20,21,22}. 

The paper is organised as follows. The relevant properties of QED gauge fixed 
in the 't~Hooft-Veltman gauge are discussed in section two with the details
of the full renormalization given in section three. Concluding comments are
provided in section four.  

\sect{Background.} 
First, we introduce the 't~Hooft-Veltman gauge in QED, \cite{16}, and the 
notation and conventions we will use. The key ingredient is the gauge fixing 
functional, ${\cal F}[A_\mu]$, which slots into the conventional path integral 
formalism for constructing a quantized gauge theory. Here $A_\mu$ is the photon
field. We take
\begin{equation}
{\cal F}[A_\mu] ~=~ \partial^\mu A_\mu ~+~ \half \xi A^\mu A_\mu
\end{equation}
which is clearly nonlinear where for the moment $\xi$ is a parameter and 
$\alpha$ is the gauge fixing parameter. Clearly when $\xi$~$=$~$0$ one recovers
the usual linear gauge fixing functional whence $\alpha$ becomes equivalent to 
the gauge parameter of those gauges. In other studies of the 't~Hooft-Veltman 
gauge, however, $\xi$ was invariably fixed to certain numerical values such as 
$1$ or $2$. We leave it as a free parameter here and given that eventually it 
will appear with the triple and quartic photon self-interactions we will regard
it as a coupling constant which will run. It is not to be confused with the 
usual gauge coupling constant, $e$, which is present in the covariant 
derivative when electrons are present. Therefore we are in effect working with 
a two coupling theory. Though in the absence of electrons, whilst photon 
self-interactions are present the field theory is effectively a free theory of 
photons since the physics cannot be altered by the gauge fixing. This feature 
ought to emerge in the computations. Hence the full Lagrangian for $\Nf$ 
massless electrons in the 't~Hooft-Veltman gauge is, \cite{16},  
\begin{equation}
L ~=~ -~ \frac{1}{4} F^{\mu\nu} F_{\mu\nu} ~+~ 
\bar{c} \partial^\mu \partial_\mu c ~+~ \xi \bar{c} A^\mu \partial_\mu c ~+~
b \left( \partial^\mu A_\mu + \frac{1}{2} \xi A^\mu A_\mu \right) ~+~ 
\frac{1}{2} \alpha b^2 ~+~ i \bar{\psi} \Dslash \psi  
\label{lagthvoff} 
\end{equation}  
where $\psi$ is the electron field, $c$ and $\bar{c}$ are the Faddeev-Popov 
ghosts emerging from the path integral formalism and $b$ is the 
Nakanishi-Lautrup auxiliary field which arises in the off-shell BRST formalism.
Eliminating it by its equation of motion produces the Lagrangian in the form we
will treat it
\begin{equation}
L ~=~ -~ \frac{1}{4} F^{\mu\nu} F_{\mu\nu} ~+~ 
\bar{c} \partial^\mu \partial_\mu c ~+~ \xi \bar{c} A^\mu \partial_\mu c ~-~
\frac{1}{2\alpha} \left( \partial^\mu A_\mu + \frac{1}{2} \xi A^\mu A_\mu 
\right)^2 ~+~ i \bar{\psi} \Dslash \psi ~. 
\label{lagthv} 
\end{equation}  
Ordinarily in a linear covariant gauge in QED one drops the Faddeev-Popov
ghosts from (\ref{lagthv}) when $\xi$~$=$~$0$ since they do not couple to 
photons or electrons. For $\xi$~$\neq$~$0$ this is not possible and they are 
not only present but play a key role in the full renormalization of the theory.
The covariant derivative, $D_\mu$, is defined by
\begin{equation}  
D_\mu ~=~ \partial_\mu ~+~ i e A_\mu ~. 
\end{equation}  
Unlike the Curci-Ferrari gauge and the MAG in QCD, there is no quartic ghost
self-interaction which is due to the absence of a colour index on the ghost
fields meaning that $c(x)c(x)$~$=$~$0$ due to their anticommuting property. The
Feynman rules for (\ref{lagthv}) are straightforward to derive but the 
interested reader can view them in \cite{23}. By construction (\ref{lagthvoff})
is invariant under the BRST symmetry, \cite{18},
\begin{equation}
\delta A_\mu ~=~ -~ \partial_\mu c ~~,~~ \delta c ~=~ 0 ~~,~~ 
\delta \bar{c} ~=~ b ~~,~~ \delta b ~=~ 0 ~~,~~ \delta \psi ~=~ i e c \psi  
\end{equation}
which is clearly nilpotent. 

\sect{Renormalization.} 
We turn now to the details of our three loop $\MSbar$ renormalization. If we
regard the fields and parameters of (\ref{lagthv}) as bare then to renormalize
(\ref{lagthv}) we introduce renormalized fields and variables by
\begin{eqnarray} 
A^\mu_{\mbox{\footnotesize{o}}} &=& \sqrt{Z_A} \, A^\mu ~~,~~ 
c_{\mbox{\footnotesize{o}}} ~=~ \sqrt{Z_c} \, c ~~,~~ 
\bar{c}_{\mbox{\footnotesize{o}}} ~=~ \sqrt{Z_c} \, \bar{c} ~~,~~ 
\psi_{\mbox{\footnotesize{o}}} ~=~ \sqrt{Z_\psi} \psi ~, \nonumber \\  
e_{\mbox{\footnotesize{o}}} &=& \mu^\epsilon Z_e \, e ~~,~~ 
\xi_{\mbox{\footnotesize{o}}} ~=~ \mu^\epsilon Z_\xi \, \xi ~~,~~ 
\alpha_{\mbox{\footnotesize{o}}} ~=~ Z^{-1}_\alpha Z_A \, \alpha 
\label{rencondef} 
\end{eqnarray} 
where the subscript ${}_{\mbox{\footnotesize{o}}}$ denotes a bare quantity. We 
have also chosen to follow the same convention as \cite{14} for the definition 
of the renormalization of the gauge parameter $\alpha$. With this convention it
is the combination $Z_A Z_\alpha^{-1}$ which is unity in the linear covariant 
gauge after renormalization. As we will use dimensional regularization in 
$d$~$=$~$4$~$-$~$2\epsilon$ dimensions with $\epsilon$ as the regularizing
parameter, the renormalization scale $\mu$ has been introduced to ensure that
both renormalized couplings $e$ and $\xi$ remain dimensionless in 
$d$-dimensions. In principle all the renormalization constants will be 
functions of both (renormalized) couplings. However, it transpires that both
the $\beta$-function of $e$ and the photon anomalous dimension are independent 
of $\xi$. Therefore, the information determined from the $\MSbar$ 
renormalization will be encoded in the renormalization group functions. The
$\beta$-functions are given by
\begin{eqnarray}
\mu \frac{\partial a}{\partial \mu} &=& \beta_a(a) ~=~ \frac{1}{2} (d-4) a ~-~
2 a \beta_a(a) \frac{\partial~}{\partial a} \ln Z_e \nonumber \\  
\mu \frac{\partial z}{\partial \mu} &=& \beta_z(a,z) ~=~ 
\frac{1}{2} (d-4) z ~-~ 2 z \beta_a(a) \frac{\partial~}{\partial a} 
\ln Z_\xi ~-~ 2 z \beta_z(a,z) \frac{\partial~}{\partial z} \ln Z_\xi ~. 
\label{betadef}
\end{eqnarray} 
The anomalous dimensions are defined in the usual way by, (see, for example,
\cite{24}), 
\begin{equation}
\gamma_A(a) ~=~ \frac{\partial \ln Z_A}{\partial \ln \mu} ~~,~~
\gamma_\alpha(a,z) ~=~ \frac{\partial \ln \alpha}{\partial \ln \mu} ~~,~~
\gamma_c(a,z) ~=~ \frac{\partial \ln Z_c}{\partial \ln \mu} ~~,~~ 
\gamma_\psi(a,z) ~=~ \frac{\partial \ln Z_\psi}{\partial \ln \mu} 
\label{basgamdef} 
\end{equation} 
from which it is straightforward to deduce 
\begin{eqnarray}
\gamma_A(a) &=& \beta_a(a) \frac{\partial~}{\partial a} \ln Z_A ~+~ 
\beta_z(a,z) \frac{\partial~}{\partial z} \ln Z_A ~+~ \alpha 
\gamma_\alpha(a,z) \frac{\partial~}{\partial \alpha} \ln Z_A \nonumber \\
\gamma_\alpha(a,z) &=& \left[ \beta_a(a) \frac{\partial~}{\partial a} 
\ln Z_\alpha ~+~ \beta_z(a) \frac{\partial~}{\partial z} \ln Z_\alpha 
~-~ \gamma_A(a) \right] \left[ 1 ~-~ \alpha 
\frac{\partial~}{\partial \alpha} \ln Z_\alpha \right]^{-1} \nonumber \\ 
\gamma_c(a,z) &=& \beta_a(a) \frac{\partial~}{\partial a} \ln Z_c ~+~ 
\beta_z(a,z) \frac{\partial~}{\partial z} \ln Z_c ~+~ \alpha 
\gamma_\alpha(a,z) \frac{\partial~}{\partial \alpha} \ln Z_c \nonumber \\
\gamma_\psi(a,z) &=& \beta_a(a) \frac{\partial~}{\partial a} \ln Z_\psi ~+~ 
\beta_z(a,z) \frac{\partial~}{\partial z} \ln Z_\psi ~+~ \alpha 
\gamma_\alpha(a,z) \frac{\partial~}{\partial \alpha} \ln Z_\psi 
\label{gamdef} 
\end{eqnarray} 
in terms of the renormalization constants where we have set 
$a$~$=$~$e^2/(16\pi^2)$ and $z$~$=$~$\xi^2/(16\pi^2)$. In these definitions, 
(\ref{gamdef}), we have not assumed that $\gamma_\alpha(a,z)$~$=$~$1$ which is 
the case in a linear covariant gauge. Clearly the wave function 
renormalizations will also depend on $\alpha$. Moreover in these definitions 
where our explicit renormalization constants are clearly independent of one of 
the coupling constants or gauge parameter, we have included this property in 
the derivation of (\ref{betadef}) and (\ref{gamdef}). For instance, $Z_e$ turns
out to independent of both $a$ and $\alpha$. 

{\begin{table}[ht] 
\begin{center} 
\begin{tabular}{|c||c|c|c|c|c|} 
\hline 
Green's function & One loop & Two loop & Three loop & Total \\ 
\hline 
$ A_\mu \, A_\nu$ & $3$ & $18$ & $~254$ & $~275$ \\ 
$ c \, {\bar c}$ & $1$ & $~\,6$ & $~~\,78$ & $~~\,85$ \\ 
$ \psi \, {\bar \psi}$ & $1$ & $~\,6$ & $~~\,78$ & $~~\,85$ \\ 
$ A_\mu \, {\bar c} \, c$ & $2$ & $33$ & $~688$ & $~723$ \\ 
$ A_\mu \, {\bar \psi \, \psi}$ & $2$ & $33$ & $~688$ & $~723$ \\ 
\hline 
Total & $9$ & $96$ & $1786$ & $1891$ \\ 
\hline 
\end{tabular} 
\end{center} 
\begin{center} 
{Table 1. Number of Feynman diagrams for the renormalization of each Green's 
function.} 
\end{center} 
\end{table}}  

The full three loop renormalization is performed for the massless case using
the {\sc Mincer} algorithm, \cite{24,25}, written in the symbolic manipulation
language {\sc Form}, \cite{26}. The Feynman diagrams are generated 
automatically with the {\sc Qgraf} package, \cite{27}, before being converted 
into {\sc Form} input notation by converter routines. The number of Feynman 
diagrams we compute at each loop order for the set of Green's functions we need
to consider to render (\ref{lagthv}) finite are given in Table $1$. The wave 
function renormalizations are deduced from the photon, Faddeev-Popov and 
electron $2$-point functions whilst the $3$-point functions determine the 
coupling constant renormalizations. For the latter given that the coupling 
constant renormalizations are gauge independent we have a strong check on the 
wave function calculations. Also for these, to apply the {\sc Mincer} algorithm
an external momentum has to be nullified. This is because {\sc Mincer} computes
massless $2$-point functions up to the finite part at three loops. The
extraction of each of the renormalization constants is found by applying the
approach of \cite{28} for automatic Feynman diagram calculations. The $2$ or 
$3$-point functions are computed as a function of the bare parameters. Then 
these are replaced by the renormalized variables from (\ref{rencondef}) and the
undetermined renormalization constant for that $2$ or $3$-point function chosen
so as to absorb the infinities which remain. The latter appear as poles in 
$\epsilon$ and are absorbed into the renormalization constants with the usual 
$\MSbar$ definition. Prior to presenting the results of our labours we note 
that one main check is that the double pole in $\epsilon$ at two loops and the 
double and triple poles at three loops for any renormalization constant are 
predetermined by the structure of the renormalization group equation. In the 
expressions we present for the anomalous dimensions and $\beta$-functions all 
the renormalization constants passed this test. 

Hence, the complete set of three loop $\MSbar$ renormalization group functions 
are
\begin{eqnarray}
\beta_a(a) &=& \frac{1}{2}[d-4] a ~+~ \frac{4}{3} \Nf a^2 ~+~ 4 \Nf a^3 ~-~ 
\left[ \frac{44}{9} \Nf^2 + 2 \Nf \right] a^4 ~+~ O(a^5) \nonumber \\  
\beta_z(a,z) &=& \frac{1}{2}[d-4] z ~+~ 
\frac{4}{3} \Nf z a ~+~ 4 \Nf z a^2 ~-~ \left[ \frac{44}{9} \Nf^2 + 
2 \Nf \right] z a^3 ~+~ O(z a^4) \nonumber \\  
\gamma_A(a) &=& \frac{4}{3} \Nf a ~+~ 4 \Nf a^2 ~-~ \left[ \frac{44}{9} \Nf^2 
+ 2 \Nf \right] a^3 ~+~ O(a^4) \nonumber \\ 
\gamma_\alpha(a,z) &=& -~ \frac{4}{3} \Nf a ~-~ 
[ 2 \alpha^2 - 3 \alpha + 3 ] \frac{z}{2\alpha} ~-~ 4 \Nf a^2 ~-~ 
[ 5 \alpha - 16 ] \frac{\Nf z a}{3\alpha} \nonumber \\
&& -~ [ 21 \alpha^3 - 20 \alpha^2 - 29 \alpha + 60 ] 
\frac{z^2}{16\alpha} ~+~ \left[ \frac{44}{9} \Nf^2 + 2 \Nf \right] a^3
\nonumber \\
&& -~ [ ( 140 \alpha - 240 ) \Nf^2 + ( 1215 \alpha - 1296 \zeta(3) \alpha 
- 6210 + 5184 \zeta(3) ) \Nf ] \frac{z a^2}{54\alpha} \nonumber \\
&& -~ [ 30 \alpha^2 - 61 \alpha - 31 ] \frac{\Nf z^2 a}{8\alpha} 
\nonumber \\
&& -~ [ 264 \zeta(3) \alpha^4 + 370 \alpha^4 - 48 \zeta(3) \alpha^3 
- 116 \alpha^3 - 1008 \zeta(3) \alpha^2 \nonumber \\
&& ~~~~-~ 1093 \alpha^2 + 432 \zeta(3) \alpha + 1778 \alpha + 360 \zeta(3) 
- 367 ] \frac{z^3}{128\alpha} ~+~ O(z^n a^{4-n}) \nonumber \\  
\gamma_c(a,z) &=& \frac{1}{4} [ 3 - \alpha ] z ~-~ \frac{5}{6} \Nf
z a ~-~ \frac{1}{32} [ 5 \alpha^2 - 16 \alpha - 5 ] z^2 \nonumber \\
&& -~ [ 140 \Nf^2 + ( 1215 - 1296 \zeta(3) ) \Nf ] \frac{z a^2}{108} ~+~
[ 13 - 17 \alpha ] \frac{\Nf z^2 a}{16} \nonumber \\
&& -~ [ 111 \alpha^3 - 48 \zeta(3) \alpha^2 - 28 \alpha^2 - 192 \zeta(3) \alpha
- 273 \alpha - 144 \zeta(3) + 410 ] \frac{z^3}{256} \nonumber \\
&& +~ O(z^n a^{4-n}) \nonumber \\ 
\gamma_\psi(a,z) &=& \alpha a ~-~ [ 4 \Nf + 3 ] \frac{a^2}{2} ~-~
[ 3 \alpha^2 - 4 \alpha + 1 ] \frac{z a}{4} ~+~ 
[ 40 \Nf^2 + 54 \Nf + 27 ] \frac{a^3}{18} \nonumber \\
&& +~ [ ( 9 \alpha - 20 ) \Nf - 8 \alpha^3 + 24 \zeta(3) \alpha - 32 \alpha
- 72 \zeta(3) + 72 ] \frac{z a^2}{4} \nonumber \\
&& -~ [ 336 \zeta(3) \alpha^3 - 323 \alpha^3 - 432 \zeta(3) \alpha^2
+ 480 \alpha^2 + 48 \zeta(3) \alpha \nonumber \\
&& ~~~~-~ 197 \alpha + 432 \zeta(3) - 344 ] \frac{z^2 a}{64} ~+~ 
O(z^n a^{4-n}) ~. 
\end{eqnarray}
where the formal order symbol $O(z^n a^{l-n})$ means all appropriate possible 
combinations of the coupling constants $a$ and $z$ at the $l$th loop. In both 
$\beta$-functions the $d$-dimensional dependence has been retained as an
indication of our conventions in deriving the renormalization group functions
as well as for the reader interested in constructing the original 
renormalization constants from the differential equations of (\ref{betadef}) 
and (\ref{gamdef}).  

Aside from the internal checks based on consistency with the renormalization
group equation there are additional checks on these results. First, both
$\beta$-functions correctly emerge as $\alpha$ independent and $\beta_a(a)$ is
in total agreement with the original linear covariant gauge result of
\cite{19,20,21,22}. Also the photon anomalous dimension is proportional to 
$\beta_a(a)$ as required by the Ward identity. Further, in the limit 
$z$~$\rightarrow$~$0$ the anomalous dimensions agree with those of the linear 
covariant gauge fixing, \cite{28}. Finally, one must recover a theory of free 
photons when the electron interaction is switched off via 
$a$~$\rightarrow$~$0$. Clearly, $\beta_z(0,z)$~$=$~$0$ which corresponds to a 
free field theory when $\xi$~$\neq$~$0$ even though neither 
$\gamma_\alpha(0,z)$, $\gamma_c(0,z)$ nor $\gamma_\psi(0,z)$ are zero. An 
additional comment on our results is that $\gamma_\psi(a,0)$ has clearly only 
$\alpha$ dependence at one loop. This was originally observed in \cite{29}, 
where it was claimed that the only $\alpha$ dependence of $\gamma_\psi(a,0)$ 
was at one loop. Clearly in a nonlinear covariant gauge there is $\alpha$ 
dependence beyond one loop which is not unexpected. Finally, in relation to the
renormalization of the gauge parameter in other conventions, we record the sum 
of $\gamma_A(a)$ and $\gamma_\alpha(a,z)$ is  
\begin{eqnarray}
\gamma_A(a) ~+~ \gamma_\alpha(a,z) &=& -~ [ 2 \alpha^2 - 3 \alpha + 3 ] 
\frac{z}{2\alpha} ~-~ [ 5 \alpha - 16 ] \frac{\Nf z a}{3\alpha}
\nonumber \\
&& -~ [ 21 \alpha^3 - 20 \alpha^2 - 29 \alpha + 60 ] \frac{z^2}{16\alpha} ~-~ 
[ 30 \alpha^2 - 61 \alpha - 31 ] \frac{\Nf z^2 a}{8\alpha} \nonumber \\
&& -~ [ ( 140 \alpha - 240 ) \Nf^2 + ( 1215 \alpha - 1296 \zeta(3) \alpha
- 6210 + 5184 \zeta(3) ) \Nf ] \frac{z a^2}{54\alpha} \nonumber \\
&& -~ [ 264 \zeta(3) \alpha^4 + 370 \alpha^4 - 48 \zeta(3) \alpha^3 
- 116 \alpha^3 - 1008 \zeta(3) \alpha^2 \nonumber \\
&& ~~~~-~ 1093 \alpha^2 + 432 \zeta(3) \alpha + 1778 \alpha + 360 \zeta(3) 
- 367 ] \frac{z^3}{128\alpha} \nonumber \\
&& +~ O(z^n a^{4-n})  
\label{alpren}
\end{eqnarray}
which is clearly non-zero for $z$~$\neq$~$0$. Moreover, like the MAG, (see, for
instance, \cite{15}), the corresponding anomalous dimension is also singular in
the $\alpha$~$\rightarrow$~$0$ limit, though similarly the remaining 
renormalization group functions, including the $\beta$-functions, are finite in
this limit. For $\gamma_A(a)$, $\gamma_c(a,z)$ and $\gamma_\psi(a,z)$ this is 
primarily because in (\ref{basgamdef}) and (\ref{gamdef}), the term involving 
$\gamma_\alpha(a,z)$ is multiplied by $\alpha$ {\em and} $Z_A$, $Z_c$ and 
$Z_\psi$ themselves are non-singular at $\alpha$~$=$~$0$.  

Finally, it might be tempting to try and remove the $\alpha$~$=$~$0$
singularity in $\gamma_\alpha(a,z)$ by a suitable coupling constant 
redefinition. Whilst this would produce renormalization group functions
analytic in $\alpha$, one must be careful in ensuring that the original theory 
is retained. For instance, to remove the $1/\alpha$ terms in 
$\gamma_\alpha(a,z)$ the least one must do is to redefine $z$ by a factor 
proportional to $\alpha$. In the simplest case, setting $z$~$=$~$\alpha\bar{z}$
one would formally have non-singular renormalization group functions. However, 
in this instance returning to (\ref{lagthv}) in the absence of electrons the 
Landau gauge Lagrangian would then describe self-interacting photons with 
non-interacting ghosts. This is not consistent with the notion that without 
electrons the photon is a free field. Therefore given this, avoiding what might
be perceived to be a problem in a renormalization group function, which has no 
physical interpretation, in order to render it analytic, has a significant 
affect on the nature of the original theory. In other words whilst it might 
seem unnatural to have a theory with couplings which are singular as 
$\alpha$~$\rightarrow$~$0$ leading to a singular anomalous dimension, the 
nature of the theory remains consistent.

\sect{Discussion.} 
We have completed the full three loop renormalization of QED in the nonlinear
't~Hooft-Veltman gauge. This extends the one loop calculations of \cite{17,18}.
Whilst the authors of \cite{17,18} were the first to observe that the 
longitudinal part of the photon is renormalized unlike in a linear gauge, we 
have carried out a slightly more general analysis by allowing for a covariant 
gauge parameter $\alpha$ and the inclusion of an additional coupling $\xi$ in 
order to track the loop calculation in a similar way to the usual coupling 
constant $a$. The coupling $\xi$ was not initially fixed to a specific value.
Consequently the singular renormalization of the gauge parameter emerges. 
Whilst this is not a new feature of a nonlinear gauge fixing, since the MAG of 
QCD has the same property, it does not disrupt either the renormalizabilty of 
the theory, \cite{16}, or the evaluation of gauge independent quantities such 
as the $\beta$-functions. This is primarily because although the gauge in one 
sense is only defined in the $\alpha$~$\rightarrow$~$0$ limit, (\ref{lagthv}), 
the anomalous dimension of $\alpha$, (\ref{alpren}), has no physical meaning or
interpretation.  

\vspace{0.5cm}
\noindent
{\bf Acknowledgements.} The author thanks Dr D. Dudal and Dr J. Babington for 
valuable discussions.

\end{document}